\documentclass{elsart}

\usepackage{graphicx}

\begin{document}

\begin{frontmatter}
\title{Studies of R$_2$Ti$_2$O$_7$ (R=Gd and Yb);\\ new results}
\author[CEAG]{P. Dalmas de R\'eotier},
\author[CEAG]{V. Glazkov},
\author[CEAG]{C. Marin},
\author[CEAG]{A. Yaouanc},
\author[IRI]{P.C.M. Gubbens},
\author[IRI]{S. Sakarya},
\author[CEAS]{P. Bonville},
\author[PSI]{A. Amato },
\author[PSI]{C. Baines},
\and \author[RAL]{P.J.C. King}

\address[CEAG]{CEA/DSM, D\'epartement de Recherche 
Fondamentale sur la Mati\`ere Condens\'ee,
F-38054 Grenoble Cedex 9, France}
\address[IRI]{Department of Radiation, Radionuclides \& Reactors,
Delft University of Technology, 2629 JB Delft, The Netherlands} 
\address[CEAS]{CEA/DSM, D\'epartement de Recherche sur 
l'Etat Condens\'e, les  Atomes et les Mol\'ecules,
F-91191 Gif sur Yvette, France}
\address[PSI]{Laboratory for Muon-Spin Spectroscopy, Paul Scherrer Institute, 
5232 Villigen-PSI, Switzerland}
\address[RAL]{ISIS Facility, Rutherford Appleton Laboratory, Chilton,
Oxfordshire OX11 0QX, United Kingdom}
\begin{abstract}

Specific heat and muon spin rotation and relaxation data are presented
for two geometrically frustrated systems: Gd$_2$Ti$_2$O$_7$ which
antiferromagnetically orders and 
Yb$_2$Ti$_2$O$_7$ which presents dynamical short range correlations
at low temperature. The muon data help to characterize the spin dynamics of 
these two compounds.

\end{abstract}
\end{frontmatter}

Geometrically derived magnetic frustration arises when the spatial 
arrangement of the spins is such that it prevents the simultaneous 
minimization of all interaction energies \cite{Ramirez01}. 
Compounds of interest for the investigation of frustration are, 
for example, the crystallographically ordered pyrochlore structure compounds 
R$_2$Ti$_2$O$_7$ where the rare earth ions (R) form a sub-lattice of
corner sharing tetrahedra.

We shall focus on two pyrochlore compounds 
Gd$_2$Ti$_2$O$_7$ and Yb$_2$Ti$_2$O$_7$. Reports of experimental
data concerning these two 
systems were already published by some of us, see
Refs. \cite{Yaouanc05,Dalmas05,Hodges02,Bonville04}.
A summary of the muon spin rotation and relaxation ($\mu$SR) data 
has been recently provided \cite{Dalmas04}. Here we report
unpublished data obtained from specific heat and $\mu$SR measurements. 
The latter measurements were either performed at the ISIS $\mu$SR facility
in UK or at the Swiss Muon Source of the Paul Scherrer Institute
in Switzerland.
We stress that the specific heat measurements are essential to 
test the sample quality in geometrically frustrated systems.
For example, in the case of Gd$_2$Ti$_2$O$_7$, the heat 
treatment affects the specific heat data in two ways. First it notably 
increases the intensity of the signal and second it slightly decreases the 
temperature of the two peaks corresponding to the magnetic transitions. 
As far as $\mu$SR spectra are concerned, the 
values for the measured local fields are noticeably changed by the
heat treatment.

\begin{figure}
\includegraphics[scale=0.8]{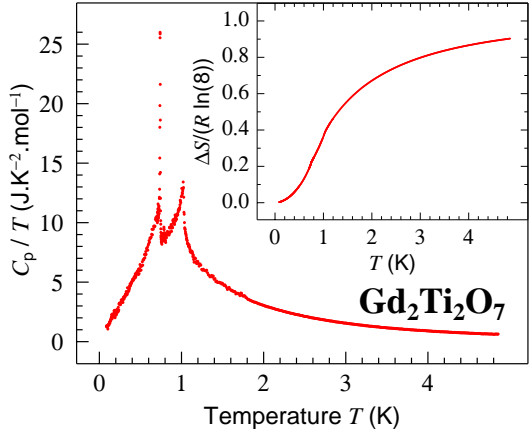}
\caption{Thermal dependence of the specific heat divided by the temperature 
measured for a crystal of
Gd$_2$Ti$_2$O$_7$. The data are presented per mole of Gd.
The maxima of the specific heat at $T_{\rm c1} = $ 1.02~K
and $T_{\rm c2}= $ 0.74~K  are the signatures of the magnetic phase 
transitions. The insert displays the computed entropy 
variation as a function of the temperature between $\simeq$ 85 mK and 5 K.}
\label{fig_Cp_Gd}
\end{figure}

\begin{figure}
\includegraphics[scale=0.8]{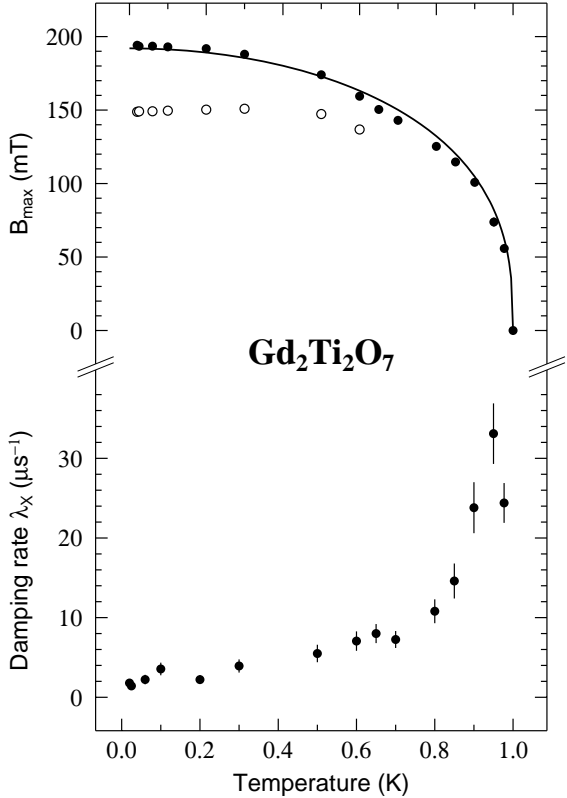}
\caption{Temperature dependence of $B_{\rm max}$ measured for Gd$_2$Ti$_2$O$_7$
(two fields per temperature point can be resolved
up to 0.6 K and only one field at higher temperatures) and the associated exponential
damping rate $\lambda_X$ for the higher $B_{\rm max}$ value. The damping rate
is negligible 
for the smaller $B_{\rm max}$. The line is the result of a fit for 
the larger field
with $B_{\rm max}(T)\propto [1- (T/T_{\rm c1})^\alpha]^\beta$, setting $\alpha = 2$ and
$\beta = 0.365$ as expected for a 3D Heisenberg antiferromagnet characterized 
by a second order phase transition at $T= T_{\rm c1}$. The fit is numerically
reasonable and physically justified since $B_{\rm max}$ is expected to track 
the order parameter.  
The contact field term, which might have a temperature
dependence different from that of the order parameter, should indeed be 
negligible for an insulating compound such as Gd$_2$Ti$_2$O$_7$.}
\label{fig_fre_lambdax_Gd}
\end{figure}

In a first step we consider Gd$_2$Ti$_2$O$_7$. Figure~\ref{fig_Cp_Gd} shows 
the specific heat measurements and the variation of entropy which is deduced
from these data. The entropy variation up to $\simeq$ 5 K is only 90 \%
of its expected value for a spin 7/2 system. This missing entropy is probably 
related to the persistent spin dynamics observed in this system deep in the
ordered state \cite{Yaouanc05}.
The specific heat structure observed at $T_{\rm c1} \simeq$ 1.02 K for 
Gd$_2$Ti$_2$O$_7$ 
corresponds to a long-range magnetic ordering of the Gd magnetic moments as 
shown by the spontaneous muon spin precession detected below $T_{\rm c1}$ in 
zero-field. In fact, we observe two 
Bessel-like oscillations, each characterized by the maximum field at the 
muon site,
$B_{\rm max}$; see Fig.~\ref{fig_fre_lambdax_Gd}.
Note that the transition at $T_{\rm c2}\simeq$ 0.74 K is hardly visible in 
the temperature
dependence of $B_{\rm max}$. This is not surprising since the difference
between the magnetic structures adopted by Gd$_2$Ti$_2$O$_7$ above and below
$T_{\rm c2}$ is small: the change concerns only a quarter of the Gd$^{3+}$ 
moments which are disordered above $T_{\rm c2}$ and very weakly ordered
below \cite{Stewart04}.

\begin{figure}
\includegraphics[scale=0.8]{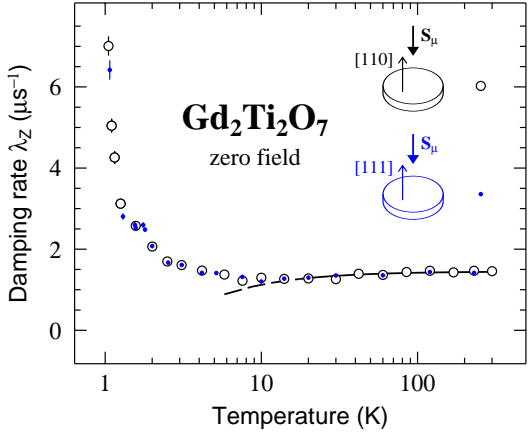}
\caption{The spin-lattice relaxation rate $\lambda_Z$ versus temperature measured in the 
paramagnetic phase for crystals of Gd$_2$Ti$_2$O$_7$. The data have been 
recorded for two orientations of the initial muon polarization, parallel to
either a [111] or [110] crystallographic axis: they show
no difference. The line is the result of a fit 
with $\lambda_Z (T) = \lambda_Z^{(ss)} (1 + T_0/T)$ where
$\lambda_Z^{(ss)}= 1.45 \, (37) \, \mu{\rm s}^{-1}$ and $T_0 = -2.3 \, (6)$ K.
The further increase of $\lambda_Z$
as the sample is cooled down below $\simeq$ 10 K   
reflects the slowing down of the magnetic fluctuations
when approaching $T_{\rm c1}$ from above. This effect is not accounted for by 
the model.}
\label{fig_lambdaZ_Gd}
\end{figure}

Although the thermal behavior of the spin-lattice relaxation, $\lambda_Z$, at 
high temperature is not directly related to the frustrated nature of the 
compounds of interest, it is still of interest. We present $\lambda_Z(T)$ 
above $T_{\rm c1}$
for  Gd$_2$Ti$_2$O$_7$ in Fig.~\ref{fig_lambdaZ_Gd}. When the temperature is
decreased from room temperature towards $T_{\rm c1}$, $\lambda_Z(T)$
first slightly decreases and reaches a local minimum around 10 K, before the 
slowing down of the spin dynamics causes a marked increase of $\lambda_Z(T)$
when approaching $T_{\rm c1}$. The initial slight decrease is the signature of 
the antiferromagnetic spin correlations which build up in Gd$_2$Ti$_2$O$_7$ 
as the system is cooled down. It is fitted to a model
in which the spin correlation function is expanded versus $(k_BT)^{-1}$
\cite{Dalmas04,Dalmas05} (see line in Fig.~\ref{fig_lambdaZ_Gd}). 

\begin{figure}
\includegraphics[scale=0.8]{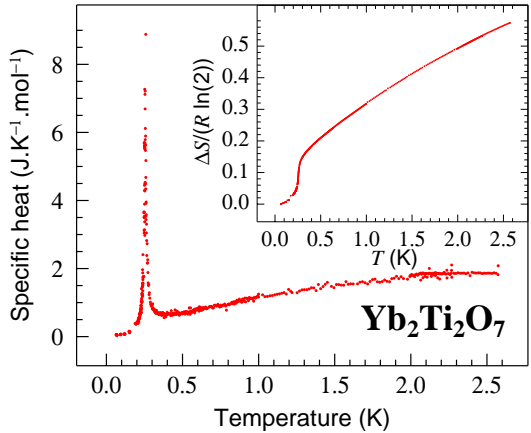}
\caption{Specific heat variation for the powder sample of 
Yb$_2$Ti$_2$O$_7$ which was used in a previously published study 
\cite{Hodges02}. The data are plotted per mole of Yb.
The maximum of the peak is around 0.25 K, which is the temperature
of a sharp change in the spin dynamics \cite{Hodges02}. The insert shows
the entropy variation deduced from the specific heat data.}
\label{fig_Cp_Yb}
\end{figure}

\begin{figure}
\includegraphics[scale=0.8]{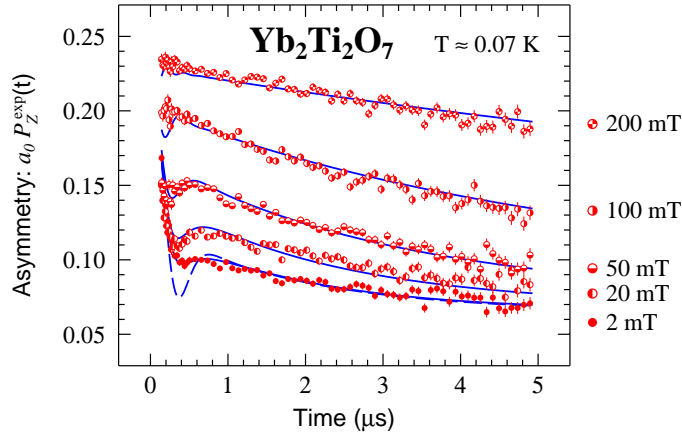}
\caption{Effect of a longitudinal magnetic field on the longitudinal 
polarization function 
$P_Z(t)$ recorded in Yb$_2$Ti$_2$O$_7$ at low temperature. The spectra are 
presented in absolute scale,
{\sl i.e.} they are not shifted vertically.  The different lines are  
results of fits described in the main text.}
\label{fig_lambdaZ_Yb}
\end{figure}

A remarkable output of most $\mu$SR measurements for geometrically frustrated
magnetic materials has been the observation of persistent spin dynamics at low
temperature. That dynamics has recently been shown to be the signature of a 
large density of states at low energy characterized by a gap linear in 
temperature \cite{Yaouanc05}. The persistent dynamics is usually recognized 
by the observation of a finite value for $\lambda_Z$. This is the case for 
Gd$_2$Ti$_2$O$_7$ \cite{Yaouanc05}. If the spin dynamics is 
particularly slow, $P_Z(t)$ is no longer stretched-like but Kubo-Toyabe-like.
This type of relaxation function was found in
Yb$_2$Ti$_2$O$_7$ \cite{Hodges02} below the temperature of the specific 
heat peak shown in Fig. \ref{fig_Cp_Yb}. The spin dynamics is then revealed
by the finite slope of $P_Z(t)$ at large time. In order to ascertain the 
interpretation of the Yb$_2$Ti$_2$O$_7$ spectra, we have investigated their 
field 
dependence. As shown in Fig.~\ref{fig_lambdaZ_Yb}, the relaxation is quenched
by a longitudinal field. 
The fits presented as full lines in this figure are done using the function 
$a_0 P_Z(t)=a_1 P_{\rm GBG}(t)+a_2$ where $P_{\rm GBG}(t)$ is the
Gaussian-broadened Gaussian function \cite{Noakes97} and the second term 
($a_2$) represents the background signal which mainly arises from muons
stopped in the silver sample holder.
A conventional Kubo-Toyabe function instead of $P_{\rm GBG}(t)$ does not 
properly account for the data. This is illustrated by the dashed line in 
Fig.~\ref{fig_lambdaZ_Yb} which is the best fit for the spectrum recorded in
a 2 mT longitudinal using the Kubo-Toyabe function: the dashed line curve 
differs notably from the 
data for $t\le $ 1 $\mu$s. It means that the muon spins 
detect a field distribution influenced by disorder. 
This probably results from the geometrical frustration of the magnetic interactions. We also note that a dynamical spin-glass model \cite{Uemura85} does not
either provide a fit of the early times data as good as the
$P_{\rm GBG}(t)$ function.

The model behind this Gaussian-broadened Gaussian function assumes that the 
muons see different environments, 
each of them being characterized by a dynamical Kubo-Toyabe function associated
with a field width $\Delta/\gamma_\mu$. To account for the slight differences
between the environments, $\Delta$ is then assumed to be Gaussian distributed 
with a mean-value $\Delta_0$ and a width $w$.
The common parameters for the five spectra presented in 
Fig.~\ref{fig_lambdaZ_Yb} are the initial asymmetries 
of the $\mu$SR signals related to the sample and the background, 
respectively $a_1$ = 0.170 and $a_2$ = 0.065, $\Delta_0 = 4.9 \, 
\mu {\rm s}^{-1}$ ($\Delta_0/\gamma_\mu$ = 5.7 mT), the correlation frequency
of the field at the muon $\nu$ = 0.85 MHz and $w/\Delta_0$ = 0.38.
Only the field values are changed. As expected from the important effect of
modest fields in the shape of
the $\mu$SR spectra, $\nu$ is found in the megahertz range. 
The remarkable result is that the fits work only if we 
take the values for the field smaller than the actual applied field. 
We have $B_{\rm meas}$ = 2.5, 
5.0, 9.2, 17.2 and 36 mT instead of $B_{\rm ext}$ = 10, 20, 50, 100 and 
200 mT respectively. This remarkably feature
has already been encountered for a Kagome-like material \cite{Uemura94}.
It has been attributed to the intermittent nature of the spin dynamics. 
In our case the ratio $B_{\rm meas}$/$B_{\rm ext}$ depends on $B_{\rm ext}$
which would imply that the applied field affects the system. We note that
$\Delta_0/\gamma_\mu$ is reduced compared to the high temperature field
width of 80 mT \cite{Yaouanc03} measured above the specific heat anomaly at 
$\simeq 0.25$~K. This reduction is consistent with the sporadic nature
of the field at the muon. More work is yet needed for an interpretation of
the magnitude of this reduction.

Financial support from the Netherlands Organization for Scientific 
Research (NWO) and the European Union (EU) is gratefully acknowledged.

\end{document}